\documentclass[11pt,a4paper]{article}
\usepackage{amsmath,amssymb}
\usepackage{epsfig,graphicx}
\newcommand{\beq}{\begin{equation}}
\newcommand{\eeq}{\end{equation}}
\newcommand\beqa{\begin{eqnarray}}
\newcommand\eeqa{\end{eqnarray}}
\newcommand\bea{\begin{array}}
\newcommand\eea{\end{array}}
\newcommand{\nn}{\nonumber}
\newcommand{\neqa}{\nonumber\end{eqnarray}}
\newcommand{\la}{\label}

\newcommand{\eq}[1]{eq.(\ref{#1})}

\newcommand{\Tr}{{\rm Tr}}

\newcommand{\Det}{{\rm Det}}

\renewcommand{\d}{\partial}
\newcommand{\OO}{{\cal O}}

\newcommand{\D}{r_{12}}

\newcommand{\bv}{\overline{{\rm v}}}

\newcommand{\vv}{{\rm v}}

\newcommand{\re}{\relax{\rm I\kern-.18em R}}

\def\su2{{SU(2)}}

\begin{document}

\begin{flushright}
LPTENS-07/01\\
hep-th/0701192
\end{flushright}

\vspace{1cm}
\begin{center}
{\Large\bf~Determinants in QCD at finite temperature} \vspace{1cm}

{\large Nikolay ~Gromov \footnote{{\bf e-mail}: gromov@thd.pnpi.spb.ru}}
\\
\vspace{4mm}
\small{\em St.Petersburg INP Gatchina, 188 300, St.Petersburg, Russia,}\\
\small{\em Laboratoire de Physique Th\'eorique de l'Ecole Normale Sup\'erieure\footnote{Unit\'e mixte du C.N.R.S. et de l' Ecole Normale Sup\'erieure, UMR 8549.}}\\
\small{\em  et l'Universit\'e Paris-VI, Paris, 75231, France}

\end{center} 
\vspace{1cm}

\begin{abstract}
We compute the functional determinant for the fluctuations around
the most general self-dual configuration with unit topological
charge for 4D $SU(2)$ Yang-Mills with one compactified direction.
This configuration is called  ``instanton with non-trivial
holonomy" or ``Kraan-van-Baal-Lee-Lu caloron". It is a
generalization of the usual instantons for the case of non-zero
temperature. We extend the earlier results of Diakonov, Gromov, Petrov and Slizovsky to
arbitrary values of parameters.

\end{abstract}
\vspace{1cm}

\section{Introduction}
Since the pioneering work of Callan, Dashen and
Gross~\cite{Callan1}, where it was proposed to approximate the QCD
path-integral by a superposition of
instantons, many authors succeeded in developing and applying the
instanton liquid model \cite{Diakonov3}. This
model has many lattice and phenomenological confirmations. An
instanton-like lumpy structure has been observed in lattice
studies using various cooling techniques
~\cite{Ilgenfritz1}. The
instanton liquid model  successfully explains the chiral symmetry
breaking~\cite{Banks1}, describes hadronic correlators and details
of hadronic structure~\cite{Shuryak2} and solves
the $U(1)_A$ problem~\cite{tHooft1}.

However, the standard instanton liquid model could not describe confinement~\cite{Chen1}.
In~\cite{DGPS} it was shown analytically that consideration of the more general
 solutions with non-trivial holonomy (KvBLL calorons)~\cite{KvB,Kraan3} leads to the
 existence of two phases with phase transition temperature $T_c\simeq 1.1\Lambda$.
In \cite{Gerhold:2006sk} a dilute
gas of KvBLL
calorons was studied in detail. This approach is also motivated
by lattice
observations~\cite{Shcheredin1}. See also recent review \cite{vanBaal:2006nx} of this activity.

\begin{figure}[t]
\centerline{ \epsfxsize=0.6\textwidth
\epsfbox{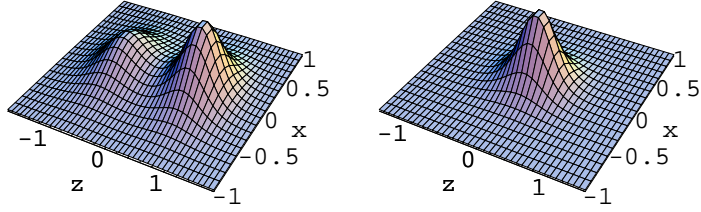}} \caption{\label{fig:adp2} The
action density of KvBLL caloron as a function of $z,x$ at
fixed $t=y=0$. At large separations $r_{12}$ the caloron is a
superposition of two BPS dyon solutions (left, $r_{12}=1.5/T$). At
small separations they merge (right, $r_{12}=0.6/T$).}
\end{figure}

The KvBLL caloron \cite{KvB} is a generalization of the BPST
instanton \cite{BPST} and Harrington-Shepard instanton ~\cite{HS}
to nontrivial holonomy. It is a self-dual gauge field
configuration, periodical in one Euclidean time direction with
period $1/T$, where $T$ is a temperature. It is characterized by
an additional gauge invariant -- holonomy or eigenvalues of the
Wilson line that goes along the time direction.  The fascinating feature of KvBLL
caloron is that it consists of two BPS dyons for $SU(2)$ gauge
group (see fig.\ref{fig:adp2}). Recently the
higher charge calorons were obtained \cite{Bruckmann:2004nu,KvBmult}.

In general, to take into account the effect of
quantum fluctuations around a classical solution
one expands the Euclidean action as follows~\cite{Gervais1}
\begin{equation}
Z_{\rm cl} = e^{-S_{\rm cl}} \int d({\rm collective\, coordinates})\cdot {\rm Jacobian}\cdot
{\rm Det'}^{-1/2}(W_{\mu \nu})\cdot {\rm Det}(-D_\mu^2),
\la{eq:oneinstanton}
\end{equation}
which is the single-pseudoparticle contribution to the partition
function. Here $D_\mu$ is a covariant derivative in the adjoint
representation, ${\rm Det'}(W_{\mu \nu})$ denotes the non-zero
mode determinant of the quadratic form of the Euclidean action,
parametrized by collective coordinates
 of the classical solution. For self-dual background fields one can show~\cite{BC} that
 ${\rm Det'}(W_{\mu \nu})={\rm Det}^4(-D_{\mu}^2)$. Thus
the determinant ${\rm Det}(-D_{\mu}^2)$ determines the weight of
the quasi-particle or the probability with which it occurs in the
partition function of the theory. The quantum determinant for the
case of zero temperature was computed by 't Hooft \cite{tHooft1}
in 70's and it still plays an important role in phenomenological
and theoretical studies of strong interaction physics. The finite
temperature generalization was performed by Gross, Pisarski and
Yaffe \cite{GPY}. They found the weight of the instanton with
a trivial holonomy i.e. with unit value of the Wilson loop
 going along the periodic Euclidean time direction. More recently
exact analytical expressions were derived for the determinants in
the fundamental and adjoint representations of the $SU(2)$ gauge
group and \textit{arbitrary} holonomy in \cite{DGPS,GS}.
Unfortunately, these expressions were extremely cumbersome and
occupied a significant amount of a hard disk space. Nevertheless,
in \cite{GS} the results of numerical evaluation were presented.
Several approximate results have recently been established for the
SU(N) case \cite{GSFermSUN}.

In this article we argue the existence of a simple relation
between determinants in the adjoint and fundamental
representations. Namely, if the determinant in the fundamental
representation is written in the form \cite{GS} (we take $T=1$ and
restore exact $T$-dependence in the last section only)
\beqa
\left.\log\Det(-\nabla^2)\right|_{T=1}&=&\left.\log\Det(-\nabla^2)\right|_{T=0}+A(\vv,\D)\\
\nn&+&V\left[P\left(\frac{2\pi-\vv}{2}\right)-\frac{\pi^2}{12}\right]+\frac{\pi\D}{2}P''\left(\frac{2\pi-\vv}{2}\right)\;,
\eeqa
where $P(\vv)=\frac{\vv^2(2\pi-\vv)^2}{12\pi^2}$ - a perturbative potential, $\vv$ - a quantity connected with holonomy (when $\vv=0$ and $\vv=2\pi$
the holonomy is trivial, see \cite{DGPS,GS} for more notations), and $\D$ is a distance between constituent dyons or $\D = \pi\rho^2 T$, where $\rho$
is an instanton size. The determinant in the adjoint representation is simply
\beqa \la{relation}
\left.\log\Det(-D^2)\right|_{T=1}&=&\left.\log\Det(-D^2)\right|_{T=0}+16 A(\vv,\D)+\log\left(1+\frac{\D\vv\bv}{2\pi}\right)\\
\nn&+&V P(\vv)+2\pi\D P''(\vv),
\eeqa
where we denote $\bv=2\pi-\vv$. This connections is in spirit of the one found by Gross, Pisarski and Yaffe. This relation provides an independent
check of the results of \cite{DGPS}. In particular the large $\D$ asymptotic, found there, can be easily rederived on the base of this relation. All
analytical and numerical results of \cite{GS} extend automatically to the isospin-1 case. The function $A(\vv,\D)$ is know with a good accuracy from
\cite{GS}.

We prove the relation (\ref{relation}) in the following way: using an exact expressions for the determinants \cite{DGPS,GS} we calculate analytically
the expansion in powers of $1/\D$ (see Appendix A), and check the relation up to the $1/\D^{10}$. Then we check the relation numerically for  several
values of $\D$ and $\vv$ with a precision $10^{-5}$. This calculation involves 3-fold integration of an expression of several Mb size which by itself
is rather nontrivial and is possible due to the numerical and analytical power of \textsf{Mathematica}.

In section II we review old results related to KvBLL caloron, which are important for the derivation.
In section III we derive the result basing on the $1/\D$ expansion and
in section IV we calculate the quantum weight of KvBLL caloron and
make more accurate the estimate for the the phase transition temperature made in \cite{DGPS}.

\section{Old results}
Before proceeding to argue the relation (\ref{relation}) between determinants in different representations let us first
remind results concerning determinants in the background of KvBLL caloron.
\subsection{Zero temperature}
When the size of KvBLL caloron $\rho$ or distance between constituent BPS dyons
$\D=\pi\rho^2 T$ is small compared to $1/T$ the caloron reduces to the usual BPST
instanton.
We recall the results by 't Hooft for isospin-$1/2$ and isospin-$1$ zero-temperature instanton determinants
 \cite{tHooft1}:
\beqa
\la{DinstFun}
\left.\log \Det(-\nabla^2)\right|_{T= 0}&\!\!\!=\!\!\! & \frac{1}{6} \log \mu\rho +\alpha(1/2) \;,\;\;\;\;\;
\alpha(1/2)=\frac{\gamma_E}{6}-\frac{17}{72}+\frac{\log\pi}{6}-\frac{\zeta'(2)}{\pi^2}\\
\la{DinstAdj}
\left.\log \Det(-D^2)\right|_{T= 0}&\!\!\!=\!\!\! & \frac{2}{3} \log \mu\rho +\alpha(1),\;\;\;\;\;
\alpha(1)=\frac{2\gamma_E}{3}- \frac{16}{9} +\frac{\log 2}{3} +  \frac{2\log (2\pi )}{3}
-\frac{4\zeta'(2)}{{\pi }^2}\;,
\eeqa
where $\mu$ is a Pauli-Villars regulator.
\subsection{Nonzero temperature, trivial holonomy}
The determinant in case of the trivial holonomy and non-zero temperature was calculated by Gross, Pisarski and Yaffe ~\cite{GPY}.
At the trivial holonomy the caloron becomes spherically symmetric.
Consequently, the resulting expressions are much
simpler. Nevertheless, even for this simpler case it has not been shown analytically that the
isospin-1 and isospin-$\frac{1}{2}$ determinants are related.

For the isospin-$\frac{1}{2}$ the result reads
\beq
\left.\log\det(-\nabla^2)\right|_{T=1}=\left.\log\det(-\nabla^2)\right|_{T=0}+A(\D)-\frac{\pi\D}{6}\la{DetHalf0}
\eeq
where $\D=\pi\rho^2$ can be interpreted as a distance between dyons when the holonomy becomes
nontrivial
(we take $T=1$).
As it was verified numerically, the isospin-1 determinant can be written in the form
\beq
\log\det(-D^2)=\left.\log\det(-D^2)\right|_{T=0}+16A(\D)+\frac{4\pi\D}{3}\la{DetOne0}\;,
\eeq
where $A(\D)$ has the following asymptotics
\beq
A(\D)=-\frac{\pi\D}{36}+\OO\left(\D^{3/2}\right)=\frac{1}{18}-\frac{\gamma_E}{6}-\frac{\pi^2}{216}-\frac{\log(\D/\pi)}{12}+\OO\left(\frac{1}{\D}\right)
\eeq
\subsection{Non-trivial holonomy, isospin-$1/2$}
The task of calculating the determinant in the background of the caloron with nontrivial holonomy is more complicated because the field configuration
has no spherical symmetry and has an additional parameter $\vv$, that is connected with the value of the holonomy (when $\vv=0,2\pi$ the holonomy
becomes trivial). In \cite{GS} an expression for the isospin-1/2 was found for all distances $\D$ and holonomies $0\leq\vv\leq 2\pi$
\beq\la{DetHalf}
\log\Det(-\nabla^2)=\left.\log\Det(-\nabla^2)\right|_{T=0}+A(\vv,\D)+\left[P\left(\frac{\bv}{2}\right)-\frac{\pi^2}{12}\right]V+P''\left(\frac{\bv}{2}\right)\frac{\pi\D}{2}
\eeq
where the function $A(\vv,\D)$ is fitted in eq.(\ref{A_fit}) and has the following large $\D$ asymptotics (we specify more terms in \eq{AallOrd} of
Appendix A)
\beqa
\la{A_as} A(\vv,\D)&=&\frac{\log (2\pi)}{6}-\frac{\vv\log\vv}{12\pi}-\frac{\bv\log\bv}{12\pi}+\frac{1}{18}
-\frac{\gamma}{6}-\frac{\pi^2}{216}-\frac{\log (\D/\pi)}{12}\\
\nn&-& \frac{1}{12\D\pi}\left(\log(\vv\bv\D^2/\pi^2)-\frac{23\pi^2}{72}+2\gamma+\frac{37}{6}\right) +\OO\left(\frac{1}{\D^2}\right)
\eeqa
and for small $\D$ it is
\beq
A(\vv,\D) =\frac{(3\vv\bv-2\pi^2)\D}{72\pi}+\OO\left(\D^{3/2}\right)\la{Aasy}
\eeq
Note that eq.(\ref{DetHalf}) is a generalization of eq.(\ref{DetHalf0})
to arbitrary values of
the holonomy.
\subsection{Non-trivial holonomy, isospin-$1$}
Isospin-$1$ or ghost determinant plays an important role since it
determines the statistical weight of the configuration.
In \cite{DGPS} an analytical expression for its large-$\D$ asymptotics was found:
\beqa
\log\Det(-D^2)&=&V\,P(\vv)+\frac{2}{3}\log\mu+\frac{3\pi-4\vv}{3\pi}\log\vv
+\frac{3\pi-4\bv}{3\pi}\log\bv+\frac{5}{3}\log(2\pi)+2\pi P''(\vv)\,\D\\
\nn&+&\frac{1}{\D}\left[\frac{1}{\vv}+\frac{1}{\bv} +\frac{23\pi}{54}-\frac{8\gamma_E}{3\pi}-\frac{74}{9\pi}
-\frac{4}{3\pi}\log\left(\frac{\vv\bar\vv\,\D^2}{\pi^2}\right)\right] +c_1+\OO\left(\frac{1}{\D^2}\right) \la{Nthisoone}
\eeqa
where
\beqa
c_1=\log 2+\frac{5}{3} \log \pi-\frac{8}{9}-2\,\gamma_E
-\frac{2\,\pi^2}{27}-\frac{4\,\zeta'(2)}{{\pi }^2}\,.
\la{c1}
\eeqa
The most nontrivial part is a constant $c_1$.
It can be easily rederived independently, using the result for isospin-$1/2$
(\ref{A_as}) and the relation (\ref{relation}).

\section{Derivation of the relation}
\begin{figure}[t]
\centerline{
\epsfxsize=0.5\textwidth
\epsfbox{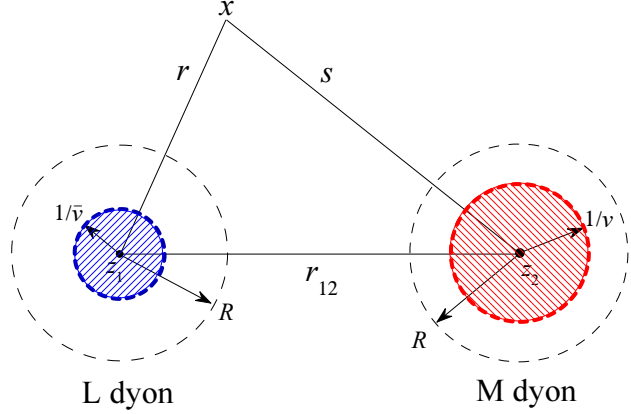}}
\caption{\label{fig:LMdyons} Three regions of integration for well separated dyons.}
\end{figure}
In this section we derive the relation (\ref{relation}) by comparing the
large-$\D$ asymptotics. We shall check this relation up to the
$10^{\rm th}$ order in $1/\D$. The method of the calculation is taken from \cite{DGPS}.

A derivative of the determinant with respect to a parameter of the background field is
\beq
\frac{\d\log\det(-D^2)}{\d\D}\equiv -\int\Tr\left(\d_{\D} A_\mu J_\mu\right)
\eeq
Here $J_\mu$ is a vacuum current related to Green function of the covariant Laplas operator in the background of KvBLL caloron. One of the results of
\cite{DGPS} and \cite{GS} is an expression for the vacuum current $J_\mu$ that is a rational function of $r,\;s,\;R=e^{r\bv},\;S=e^{s\vv},\;E_0=e^{2
i \pi x_0}$ and $\vv$, where $r,\; s$ are distances from the BPS dyons (see. fig.\ref{fig:LMdyons}), $1/\vv$ and $1/\bv$ are their core sizes.

The main point in the expansion is to divide space into tree domains: two balls of radius $R$, such that $1/\vv,1/\bv\ll R\ll\D$, surrounding the
centers of the constituent BPS dyons, and all the rest space. Then we expand the expression
in powers of $1/\D$ near each core
and integrate it over the core domains. The expression outside the cores has an exponential precision
and the only source of the $1/\D$ terms here is the nontrivial domain of integration.

The vacuum current of the isospin-$1$ can be naturally divided into tree pieces $J_\mu=J_\mu^{\rm r}+J_\mu^{\rm s}+J_\mu^{\rm m}$
(see \cite{DGPS} for notations).
Let us denote by $\frac{\d}{\d\D}\log\det^{\rm {r,s,m}}_{\rm core}$ the contributions to the
$\frac{\d}{\d\D}\log\det(-D^2)$ that comes from $J_\mu^{\rm r,s,m}$ i.e.
\beq
\frac{\d}{\d\D}\log{\det}^{\rm {r,s,m}}_{\rm core}\equiv -\int_{\rm core}\Tr\left(\d_\vv A_\mu J^{\rm r,s,m}_\mu\right)
\eeq
where integration is over two ball of the radius $R$. The total $\frac{\d}{\d\D}\log\det(-D^2)$ is a sum of these three contributions and a contribution that
comes from the integration over the rest space $\frac{\d}{\d\D}\log\det(-D^2)_{\rm far}$. In Appendix A the expansion of these
contributions in  powers of $1/\D$ is given.
One can make sure that at all orders the following equalities hold:
\beq
\frac{\d\log\Det^{\rm s}_{\rm cores}}{\d\D}=4\frac{\d A}{\d\D}-\frac{2\log\frac{\D}{R}}{3\D^2\pi}+\frac{1}{3\D}+\frac{\pi}{8\D^2}-\frac{23}{18 \D^2\pi}+(R^n\;\rm terms)
\la{ddets}
\eeq
\beq
\frac{\d\log\Det^{\rm r+m}_{\rm cores}}{\d\D}=12\frac{\d A}{\d\D}-\frac{2\log\frac{\D}{R}}{\D^2\pi}+\frac{1}{\D}+\frac{3\pi}{8\D^2}+\frac{5}{6
\D^2\pi}-\frac{2\pi}{\D(\D\vv\bv+2\pi)}+(R^n\;\rm terms)
\la{ddetrm}
\eeq
here we do not write $R^n$ terms as they all get cancelled with the similar terms
in the contribution of the 'far' region.
\beq
\frac{\d\log\Det^{\rm r+m+s}_{\rm far}}{\d\D}=\frac{8\log\frac{\D}{R}}{3\D^2\pi}+\frac{8-9\pi^2}{18\pi\D^2}+2\pi P''(\vv)+(R^n\;\rm terms)
\eeq
adding up contributions from the 'far' and the 'core' regions we have
\beq
\frac{\d\log\Det(-D^2)}{\d\D}=\d_{\D}\left(16 A+\frac{1}{3}\log\D+\log\left(2\pi+\D\vv\bv\right)+2\pi P''(\vv)\right).
\eeq
Finally we integrate it up to the small values of $\D$, where KvBLL caloron reduces to the
ordinary BPST instanton
and the determinant is known. Matching with \eq{DinstAdj} we conclude
\beq
\log\Det(-D^2)=\left.\log\Det(-D^2)\right|_{T=0}+16 A(\vv,\D)+\log\left(1+\frac{\D\vv\bv}{2\pi}\right)+2\pi P''(\vv)\D+VP(\vv) \la{Det1}
\eeq
The claim is that this answer is exact. It gives right large-$\D$ asymptotics (\ref{Nthisoone}) and is consistent with the trivial holonomy results
(\ref{DetHalf0}), (\ref{DetOne0}). Moreover, we have tested it numerically for several values of $\D$ and $\vv$ with a precision of order $10^{-5}$.
We consider this as a serious proof of the relation (\ref{Det1}).
\section{Quantum weight}
\begin{figure}[t]
\centerline{ \epsfxsize=0.5\textwidth \epsfbox{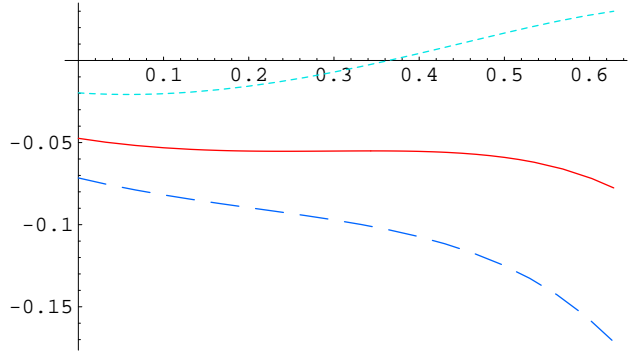}} \caption{Free energy of the caloron gas in units of $T^3V$ at $T= 1.5\Lambda$ (dotted), $T=
1.325\Lambda$ (solid) and  $T= 1.25\Lambda$ (dashed) as function of the asymptotic value of $A_4$ in units of $T$. } \la{fig_tc}
\end{figure}
In this section we renew the main result of \cite{DGPS} -- the quantum weight of KvBLL caloron.
The concept of the quantum weight is discussed in detail, for example,
in \cite{DGPS}. For a self-dual configuration it reads
\beq
{\cal Z}=\int\prod_{i=1}^p d\xi_i e^{-S_{cl}}\left(\frac{\mu}{g\sqrt{2\pi}}\right)^p J\; {\Det}^{-1}(-D^2)
\la{weight1}
\eeq
where $\xi_i$ are coordinates on the moduli space of the configuration, $g$ is a gauge coupling,
and $J$ is a measure on the moduli space. It can be expressed it terms of a metric on the moduli space
\beq
J=\sqrt{\det g_{ij}},
\eeq
in \cite{KvB} it was found that
\beq
J=8(2\pi)^8\rho^3\left(1+\frac{\D}{2\pi}\vv\bv\right).
\eeq
The total number of zero modes is 8. The associated collective coordinates are $z_\mu$ - position of KvBLL caloron center, one gauge orientation and
two angles that determine the orientation is space, combined into $\OO$, and the instanton size $\rho$. One can parameterize the moduli space by two
3D coordinates of dyons and two color orientations of the dyons. It turns our that the determinant does not depend on the color orientations.
\beq
\int\prod_{i=1}^8 d\xi_i\; J=\int d^4z\;d^4\OO\;d\rho\;\rho^3\left(1+\frac{\D}{2\pi}\vv\bv\right)16\;(2\pi)^{10} =\int
d^3z_1\;d^3z_2\;\left(1+\frac{\D}{2\pi}\vv\bv\right)\frac{1}{\D}16\;(2\pi)^{7}
\eeq
combining this with (\ref{weight1}) and (\ref{relation}) we arrive at
\beqa\nn
{\cal Z}_{\rm KvBLL}&=&\int d^3z_1 \, d^3z_2\, T^6\,C_A\left(\frac{8\pi^2}{g^2}\right)^4\!
\left(\frac{\Lambda e^{\gamma_E}}{4\pi T}\right)^{\frac{22}{3}}
\left(\frac{1}{T\D}\right)^{\frac{4}{3}}
\\\nn&\times&\exp\left[-V\, P(\vv)-16A(\vv,\D)-2\pi\,\D\, P''(\vv)\right],
\eeqa
\beq
C_A=2^8\pi^2 \exp\left(\frac{16}{9}-8\gamma_E+\frac{4\zeta'(2)}{\pi^2}+\frac{2}{3}\log 2\right).
\eeq
Surprisingly, the moduli space measure and the third term in the expression (\ref{Det1}) for $\Det(-D^2)$
exactly cancel each other.

The exact function $A(\D,\vv)$ was fitted in \cite{GS} by
\beq
\la{A_fit} A(\vv,\D)\simeq-\frac{1}{12}\log\left(1+\frac{\pi \D T}{3}\right)-\frac{\D\alpha}{216\pi(1+\D T)} +\frac{0.00302 \;\D(\alpha+9
\vv\bv/T)}{2.0488+\D^2 T^2}
\eeq
where $\alpha=18 \vv\log\frac{\vv}{T}+18\bv\log\frac{\bv}{T}-216.611 T$. This expression has a maximum absolute error $5\times 10^{-3}$.
\subsection{Estimation of the $T_c$}
We made slightly more accurate the crude estimation of the free energy
of ensemble of KvBLL calorons without taking into account an interaction made in \cite{DGPS}.
We do not repeat the details here and just give the result.

To obtain a phase transition one has to consider a gas of the calorons. The density of the calorons increases when the temperature becomes smaller.
At some critical temperature $T_c$ the density becomes sufficient to override the perturbative potential $P(\vv)$ and nontrivial values of holonomy
become preferable
 (see fig.\ref{fig_tc}). Our new estimate for $T_c$ is $1.3\Lambda$ which is sligtly bigger then in \cite{DGPS}.

\section*{Acknowledgements}
We are grateful to Dmitri Diakonov, Victor Petrov and Mikhail
Muller-Preussker and especially Sergey Slizovsky for discussions . This work was partially
supported by RSGSS-1124.2003.2, RFFI grant 06-02-16786 and French
Government PhD fellowship.

\appendix
\section{Series expansion with respect to $1/\D$}
\footnotesize
In this appendix we give results of the expansion in powers of $1/\D$.
This expressions are used to obtain \eq{ddets} and \eq{ddetrm}.

\beqa
\la{AallOrd}&&\frac{\d A(\vv,\D)}{\d\D}=-\frac{1}{24\D}+ \left[
\frac{25}{144\pi}+\frac{\gamma_E}{12\pi}-\frac{23\pi}{1728}+\frac{\log(\vv\D/\pi)}{12\pi}
\right]\frac{1}{\D^2}+\frac{1}{12\pi\D^3\vv}-\frac{1}{24\pi\D^4\vv^2}
\\&&\nn+
\left[\frac{1}{36\pi}-\frac{\pi^3}{2160}\right]\frac{1}{\D^5\vv^3}+ \left[
  -\frac{1}{48 \pi } + \frac{{\pi }^3}{576}
\right]\frac{1}{\D^6\vv^4}+ \left[
  \frac{1}{60 \pi } - \frac{73 {\pi }^3}{10800} - \frac{{\pi }^5}{2835}
\right]\frac{1}{\D^7\vv^5}
\\&&\nn+
\left[
  -\frac{1}{72 \pi } + \frac{343 {\pi }^3}{12960} + \frac{11 {\pi }^5}{3888}
\right]\frac{1}{\D^8\vv^6}+ \left[
  \frac{1}{84 \pi } - \frac{769 {\pi }^3}{8640} -
   \frac{2285 {\pi }^5}{127008} - \frac{{\pi }^7}{2016}
\right]\frac{1}{\D^9\vv^7}
\\&&\nn+
\left[
  -\frac{1}{96 \pi } + \frac{1169 {\pi }^3}{4608} +
   \frac{15025 {\pi }^5}{145152} + \frac{1111 {\pi }^7}{172800}
\right]\frac{1}{\D^{10}\vv^8}+(\vv \leftrightarrow\bv)+\OO\left(\frac{1}{\D^{11}}\right)
\eeqa

We divide $\frac{\d\log\det^{\rm r}_{\rm core}}{\d\D}$ into two parts
$\frac{\d\log\det^{\rm r1}_{\rm core}}{\d\D}$ and $\frac{\d\log\det^{\rm r2}_{\rm core}}{\d\D}$
\beqa
&&\frac{\d\log\det^{\rm r1}_{\rm core}}{\d\D}= \left[ \frac{3}{8} - \frac{{\pi }^2}{9} \right]\frac{1}{\D^2\vv}+ \left[ -\frac{3}{8} + \frac{{\pi
}^2}{9} -
   \frac{11 {\pi }^4}{1890}
\right]\frac{1}{\D^3\vv^2}+ \left[ \frac{2873}{6720} - \frac{7 {\pi }^2}{144} +
   \frac{437 {\pi }^4}{75600} + \frac{2 {\pi }^6}{2205}
\right]\frac{1}{\D^4\vv^3}
\\ \nn&&+
\left[
  - \frac{33}{70} - \frac{19 {\pi }^2}{144} +
   \frac{10693 {\pi }^4}{302400} - \frac{547 {\pi }^6}{52920}
\right]\frac{1}{\D^5\vv^4}+ \left[
  \frac{215}{448} + \frac{149 {\pi }^2}{288} -
   \frac{1651 {\pi }^4}{21600} + \frac{1321 {\pi }^6}{26460}
\right]\frac{1}{\D^6\vv^5}
\\ \nn&&+
\left[
  -\frac{353 {\pi }^2}{288} + \frac{38939 {\pi }^4}{138240} +
   \frac{8023081 {\pi }^6}{121927680} + \frac{143 {\pi }^8}{32256}
\right]\frac{1}{\D^7\vv^6}
\\ \nn&&+
\left[
  \frac{32687}{71680} + \frac{77 {\pi }^2}{32} +
   \frac{412879 {\pi }^4}{460800} + \frac{386711 {\pi }^6}{564480} +
   \frac{7049 {\pi }^8}{108000}
\right]\frac{1}{\D^8\vv^7}
\\ \nn&&+
\left[
  -\frac{153973}{337920} - \frac{407 {\pi }^2}{96} -
   \frac{4711609 {\pi }^4}{1382400} - \frac{368419 {\pi }^6}{145152} -
   \frac{2376127 {\pi }^8}{4536000} - \frac{8461 {\pi }^{10}}{1372140}
\right]\frac{1}{\D^9\vv^8}
\\ \nn&&+
\left[
  \frac{77537}{168960} + \frac{111 {\pi }^2}{16} +
   \frac{50991947 {\pi }^4}{4838400} + \frac{630755 {\pi }^6}{72576} +
   \frac{14417821 {\pi }^8}{4536000} + \frac{27623 {\pi }^{10}}{177870}
\right]\frac{1}{\D^{10}\vv^9}+(\vv \leftrightarrow\bv)+\OO\left(\frac{1}{\D^{11}}\right)
\eeqa
\beqa
&&\frac{\d\log\det^{\rm r2}_{\rm core}}{\d\D}=
\left[
\frac{5}{2 \pi } + \frac{\gamma_E}{\pi } +
   \frac{\pi }{36} + \frac{\log (\vv R/\pi)}{\pi }
\right]\frac{1}{\D^2}+
  \frac{1}{\pi }\frac{1}{\D^3\vv}-\frac{1}{2 \pi }\frac{1}{\D^4\vv^2}+
\left[ \frac{1}{3 \pi } - \frac{{\pi }^3}{180} \right]\frac{1}{\D^5\vv^3}
\\&&\nn+
\left[
  -\frac{1}{4 \pi } + \frac{{\pi }^3}{48}
\right]\frac{1}{\D^6\vv^4}+ \left[
  \frac{1}{5 \pi } - \frac{73 {\pi }^3}{900} - \frac{4 {\pi }^5}{945}
\right]\frac{1}{\D^7\vv^5}+ \left[
  -\frac{1}{6 \pi } + \frac{343 {\pi }^3}{1080} + \frac{11 {\pi }^5}{324}
\right]\frac{1}{\D^8\vv^6}
\\&&\nn+
\left[
  \frac{1}{7 \pi } - \frac{769 {\pi }^3}{720} -
   \frac{2285 {\pi }^5}{10584} - \frac{{\pi }^7}{168}
\right]\frac{1}{\D^9\vv^7}+ \left[
  -\frac{1}{8 \pi } + \frac{1169 {\pi }^3}{384} +
   \frac{15025 {\pi }^5}{12096} + \frac{1111 {\pi }^7}{14400}
\right]\frac{1}{\D^{10}\vv^8}\\
\nn&&+(\vv \leftrightarrow\bv)+\OO\left(\frac{1}{\D^{11}}\right)
\eeqa
\beqa
&&\frac{\d\log\det^{\rm s}_{\rm core}}{\d\D}=\left[\frac{1}{18 \pi } +
     \frac{\gamma_E}{3 \pi } + \frac{\pi }{108} +
\frac{\log (\vv R/\pi)}{3 \pi }\right]\frac{1}{\D^2} +  \frac{1}{3 \pi  \D^3\vv} -\frac{1}{6 \pi \D^4\vv^2} +\left[\frac{1}{9 \pi} -
     \frac{{\pi }^3}{540} \right]\frac{1}{\D^5\vv^3}
\\&&\nn+
\left[
  -\frac{1}{12 \pi } + \frac{{\pi }^3}{144}
\right]\frac{1}{\D^6\vv^4}+ \left[
  \frac{1}{15 \pi } - \frac{73 {\pi }^3}{2700} - \frac{4 {\pi }^5}{2835}
\right]\frac{1}{\D^7\vv^5}+ \left[
  -\frac{1}{18 \pi } + \frac{343 {\pi }^3}{3240} + \frac{11 {\pi }^5}{972}
\right]\frac{1}{\D^8\vv^6}
\\&&\nn+
\left[
  \frac{1}{21 \pi } - \frac{769 {\pi }^3}{2160} -
   \frac{2285 {\pi }^5}{31752} - \frac{{\pi }^7}{504}
\right]\frac{1}{\D^9\vv^7}+ \left[
  -\frac{1}{24 \pi } + \frac{1169 {\pi }^3}{1152} +
   \frac{15025 {\pi }^5}{36288} + \frac{1111 {\pi }^7}{43200}
\right]\frac{1}{\D^{10}\vv^8}\\&&+(\vv \leftrightarrow\bv) \nn+\OO\left(\frac{1}{\D^{11}}\right)
\eeqa
Analogously, we divide $\frac{\d\log\det^{\rm m}_{\rm core}}{\d\D}$ into two parts $\frac{\d\log\det^{\rm m1}_{\rm core}}{\d\D}$ and
$\frac{\d\log\det^{\rm m2}_{\rm core}}{\d\D}$. It is very convenient to extract the factor $\left(1+\frac{\D\vv\bv}{2\pi}\right)$ from the
denominator of this contribution before making an expansin
\beqa
&&-\left(1+\frac{\D\vv\bv}{2\pi}\right)\frac{\d\log\det^{\rm m1}_{\rm core}}{\d\D}=\left[\frac{11}{8} - \frac{{\pi }^2}{9}\right]\frac{1}{\D}
 -
\frac{11 {\pi }^4}{1890}\frac{1}{\D^2\vv}+\left[ \frac{353}{6720} + \frac{{\pi }^2}{16} - \frac{{\pi }^4}{25200} + \frac{2 {\pi }^6}{2205}
\right]\frac{1}{\D^3\vv^2}\\
\nn&&
+\left[
 - \frac{59}{1344} - \frac{13 {\pi }^2}{72} +
\frac{4147 {\pi }^4}{100800} - \frac{499 {\pi }^6}{52920} \right]\frac{1}{\D^4\vv^3} + \left[ \frac{19}{2240} + \frac{37\,{\pi }^2}{96} -
\frac{12421\,{\pi }^4}{302400} +
  \frac{419\,{\pi }^6}{10584}
\right]\frac{1}{\D^5\vv^4}
\\&&\nn+
\left[
\frac{-17\,{\pi }^2}{24} + \frac{115529\,{\pi }^4}{230400} +
  \frac{121033\,{\pi }^6}{2488320} + \frac{143\,{\pi }^8}{32256}
\right]\frac{1}{\D^6\vv^5}\\ \nn&&+ \left[ \frac{85\,{\pi }^2}{72} - \frac{25373531\,{\pi }^4}{11059200} -
  \frac{82760843\,{\pi }^6}{487710720} - \frac{6627049\,{\pi }^8}{96768000}
\right]\frac{1}{\D^7\vv^6}
\\&&\nn+
\left[
\frac{43}{118272} - \frac{11\,{\pi }^2}{6} -
  \frac{868243\,{\pi }^4}{345600} - \frac{4707133\,{\pi }^6}{2540160} -
  \frac{2080069\,{\pi }^8}{4536000} - \frac{8461\,{\pi }^{10}}{1372140}
  \right]\frac{1}{\D^8\vv^7}
\\&&\nn+
\left[
\frac{367}{112640} + \frac{259\,{\pi }^2}{96} +
  \frac{2555653\,{\pi }^4}{358400} + \frac{297697\,{\pi }^6}{48384} +
  \frac{95569\,{\pi }^8}{36000} + \frac{286483\,{\pi }^{10}}{1920996}\right]\frac{1}{\D^9\vv^8}+(\vv \leftrightarrow\bv)+\OO\left(\frac{1}{\D^{10}}\right)
\eeqa
\beqa
&&-\left(1+\frac{\D\vv\bv}{2\pi}\right)\frac{\d\log\det^{\rm m2}_{\rm core}}{\d\D}= \left[\frac{\pi }{18}-\frac{11}{16 \pi }\right]\frac{\vv}{\D}+
\left[ \frac{3}{16 \pi } - \frac{\pi }{18} + \frac{11 {\pi }^3}{3780}
\right]\frac{1}{\D^2}\\
\nn&&+ \left[ -\frac{2873}{13440 \pi } + \frac{7 \pi }{288} - \frac{437 {\pi }^3}{151200} - \frac{{\pi }^5}{2205} \right]\frac{1}{\D^3\vv}+\left[
\frac{33}{140 \pi } + \frac{19 \pi }{288} - \frac{10693 {\pi }^3}{604800} + \frac{547 {\pi }^5}{105840} \right]\frac{1}{\D^4\vv^2}
\\&&\nn+
\left[
-\frac{215}{896\,\pi } - \frac{149\,\pi }{576} +
  \frac{1651\,{\pi }^3}{43200} - \frac{1321\,{\pi }^5}{52920}
\right]\frac{1}{\D^5\vv^3}+ \left[ \frac{353\,\pi }{576} - \frac{38939\,{\pi }^3}{276480} -
  \frac{8023081\,{\pi }^5}{243855360} - \frac{143\,{\pi }^7}{64512}
  \right]\frac{1}{\D^6\vv^4}\\
\nn&&+
\left[
-\frac{77\,\pi }{64} + \frac{9496217\,{\pi }^3}{7372800} +
  \frac{4253821\,{\pi }^5}{36126720} + \frac{1008007\,{\pi }^7}{27648000}
\right]\frac{1}{\D^7\vv^5}\\
\nn&&+
\left[
\frac{153973}{675840\,\pi } + \frac{407\,\pi }{192} +
  \frac{4711609\,{\pi }^3}{2764800} + \frac{368419\,{\pi }^5}{290304} +
  \frac{2376127\,{\pi }^7}{9072000} + \frac{8461\,{\pi }^9}{2744280}\right]\frac{1}{\D^8\vv^5}\\
\nn&&+
\left[
-\frac{77537}{337920\,\pi } - \frac{111\,\pi }{32} -
  \frac{50991947\,{\pi }^3}{9676800} - \frac{630755\,{\pi }^5}{145152} -
  \frac{14417821\,{\pi }^7}{9072000} - \frac{27623\,{\pi }^9}{355740}\right]\frac{1}{\D^9\vv^7}
+(\vv \leftrightarrow\bv)+\OO\left(\frac{1}{\D^{10}}\right)
\eeqa

\end{document}